\documentclass[graybox]{svmult}

\usepackage{type1cm}        
\usepackage{makeidx}         
\usepackage{graphicx}        
\usepackage{multicol}        
\usepackage[bottom]{footmisc}

\usepackage{newtxtext}       %
\usepackage{newtxmath}       

\makeindex             

\usepackage[breaklinks=true,colorlinks=true,linkcolor=blue,urlcolor=blue,citecolor=blue]{hyperref}
\usepackage{cite}

\newcommand{\AswapB}{\text{A}\longleftrightarrow\text{B}}

\begin{document}

\title*{Mesoscale Modelling of the Tolman Length in Multi-component Systems}

\author{Matteo Lulli, Luca Biferale, Giacomo Falcucci, Mauro Sbragaglia and Xiaowen Shan}
\institute{Matteo Lulli, Xiaowen Shan \at Department of Mechanics and Aerospace Engineering, Southern University of Science and Technology, Shenzhen, Guangdong 518055, China, \email{lulli@sustech.edu.cn, shanxw@sustech.edu.cn}
\and Luca Biferale, Mauro Sbragaglia \at Department of Physics \& INFN, University of Rome ``Tor Vergata'', Via della Ricerca Scientifica 1, 00133, Rome, Italy \email{biferale@roma2.infn.it, sbragaglia@roma2.infn.it}
\and Giacomo Falcucci \at Department of Enterprise Engineering ``Mario Lucertini'', University of Rome ``Tor Vergata", Via del Politecnico 1, 00133 Rome, Italy; John A. Paulson School of Engineering and Applied Physics, {\it Harvard University},  33 Oxford Street, 02138 Cambridge, Massachusetts, USA. \email{giacomo.falcucci@uniroma2.it}}
%
%
\maketitle

\abstract{In this paper we analyze the curvature corrections to the surface tension in the context of the Shan-Chen (SC) multi-component Lattice Boltzmann method (LBM). We demonstrate that the same techniques recently applied in the context of the Shan-Chen multi-phase model can be applied to multi-component mixtures. We implement, as a new application, the calculation of the surface of tension radius $R_s$ through the minimization of the generalized surface tension $\sigma[R]$. In turn we are able to estimate the Tolman length, i.e. the first order coefficient of the curvature expansion of the surface tension $\sigma(R)$, as well as the higher order corrections, i.e. the curvature- and the Gaussian-rigidity coefficients. The SC multi-component model allows to model both fully-symmetric as well as asymmetric interactions among the components. By performing an extensive set of simulations we present a first example of tunable Tolman length in the mesoscopic model, being zero for symmetric interactions and different from zero otherwise. This result paves the way for controlling such interface properties which are paramount in presence of thermal fluctuations. All reported results can be independently reproduced through the ``idea.deploy" framework available at~\href{https://github.com/lullimat/idea.deploy}{\nolinkurl{https://github.com/lullimat/idea.deploy}}.}

\section{Introduction}\label{sec:Intro}
\begin{figure}[!t]
  \centering
  \includegraphics[scale = 0.58]{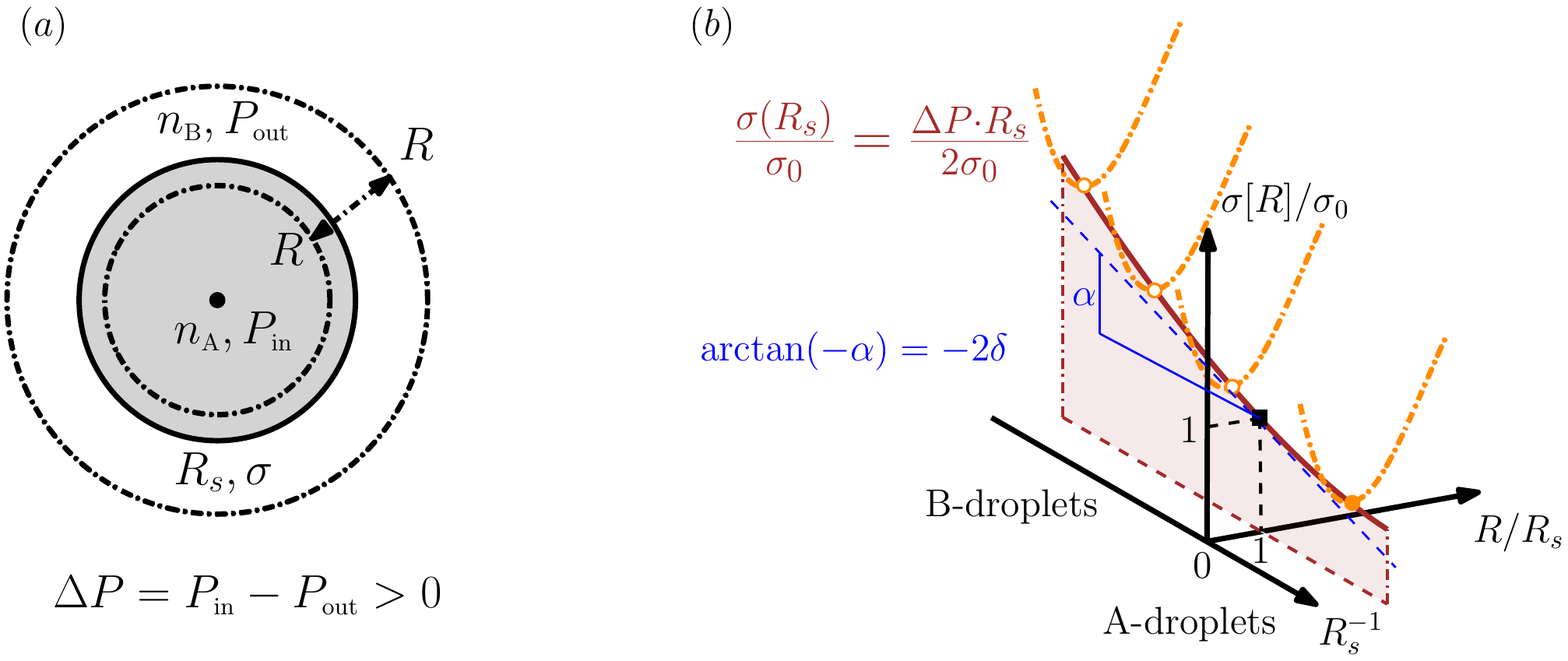}
  \caption{$(a)$: Sketch of a $n_\text{A}$-rich droplet surrounded by the $n_\text{B}$-rich fluid; the position of the \emph{arbitrary dividing surface} is indicated by $R$ while $R_s$ indicates the position of the surface of tension where the surface tension $\sigma$ acts. $(b)$: Sketch for the curvature dependence of the (normalized) surface tension, $\sigma(R_s)/\sigma_0$, reported in brown, as the locus of minima of all the generalized surface tension functions $\sigma[R]/\sigma_0$ reported in dash-dotted orange (cf. with Fig.~\ref{fig:gen_sigma_laplace}). The horizontal axes express \emph{physical} dependence on the curvature $R_s^{-1}$ and the {\emph{notional} dependence~\cite{RowlinsonWidom82}} on the (normalized) arbitrary dividing surface position $R/R_s$.}
    \label{fig:sketch}
\end{figure}

Multi-component fluids are systems where two or more components, distinguished by their chemical properties, are mutually diffused into each other. The $\text{J}$-th component can be described by the concentration $n_\text{J}$ while the flow of the mixture is characterized by a common velocity $\mathbf{u}$. In the appropriate thermodynamic conditions, i.e. temperature and pressure, for values of the concentration above a saturation threshold $n_{\text{J},\text{s}}$, it is possible for droplets of the $\text{J}$-th component to form. In particular this happens when the free-energy gain provided by the formation of a $n_\text{J}$-rich bulk region overcomes the barrier provided by the surface free-energy associated to the interface. The latter contribution is commonly described by the free-energy cost per unit area, i.e. the surface tension $\sigma$, which, in the limit of small deformations, allows to describe the mechanic response of the interface as that of an elastic membrane. Only a few configurations are mechanically stable, namely the flat and the spherical interfaces. In settings where the typical scale of the interface is such that thermal fluctuations are negligible it can be useful to adopt a simplified description of the interface as being a discontinuity point for the concentration $n_\text{J}$, i.e. going from the bulk value $n_{\text{J},\text{b}} > n_{\text{J},\text{s}}$ to a soluble value $n_{\text{J},\text{out}} < n_{\text{J},\text{s}}$ outside the bulk region. Such a discontinuity can be used to identify the \emph{surface of tension}~\cite{RowlinsonWidom82} providing a simple, yet useful, mechanical model of the interface. However, this is a somewhat idealized description since thermal fluctuations naturally induce a finite \emph{interface thickness} which opens the question of the determination of the position of the surface of tension itself. In other words, the average of the concentration profile over thermal fluctuations is a continuous curve rather than a step function. Considering a flat interface, it is possible to determine the position $z_s$ of the surface of tension by means of the pressure tensor~\cite{RowlinsonWidom82}. However, when considering spherical interfaces the determination of the surface of tension is more complicated. While in the case of a flat interface the value of $z_s$ does not explicitly enter the definition of the free-energy, for closed interfaces the curvature appears as a new control parameter with the explicit introduction of an arbitrary dividing surface~\cite{GibbsCollected1948, Buff1951, RowlinsonWidom82} whose position with respect to the center of the droplet is denoted by $R$. Since the position of such interface is arbitrary, i.e. it can either be completely inside or outside the $n_\text{J}$-rich droplet (see Fig.~\ref{fig:sketch}$(a)$), a natural request is for the free-energy to be independent, i.e. stationary, with respect to arbitrary, or notional~\cite{RowlinsonWidom82}, changes in $R$. Starting here we restrict our discussion and results to the case of two different components, hence two concentrations fields $n_\text{J}$ with $\text{J} \in \{A, B\}$. The stationarity condition reflects on the definition of a \emph{generalized} surface tension $\sigma[R]$, which assumes the shape of a convex function reaching a minimum at the surface of tension $R_s$. At the latter position the Laplace law applies in the usual form~\cite{GibbsCollected1948, Buff1951, RowlinsonWidom82, Blokhuis1992}. Indeed, it is possible to show~\cite{RowlinsonWidom82} that the stationarity of the free energy at the surface of tension $R_s$ yields
\begin{equation}\label{eq:f-statio-Rs}
    \Delta P = \frac{2\sigma[R_s]}{R_s} + \frac{\mbox{d}\sigma[R]}{\mbox{d}R}\bigg|_{R = R_s} = \frac{2\sigma(R_s)}{R_s}.
\end{equation}
By considering a generic value of $R\neq R_s$ in~\eqref{eq:f-statio-Rs} one obtains the so-called \emph{generalized} Laplace law which explicitly depends on the {notional derivative~\cite{RowlinsonWidom82}} of $\sigma[R]$.

The locus of the minima of $\sigma[R]$ identifies a physical, i.e. non-arbitrary, dependence of the surface tension $\sigma(R)$ on the droplet size at the surface of tension $R_s$, $\sigma_s = \sigma[R_s] = \sigma(R_s)$. Such a dependence was first examined, for the case of multi-phase systems, in the seminal paper by Tolman~\cite{Tolman1949} (see~\cite{Malijevsky12, Ghoufi_2016} for reviews). Similar results have been obtained for the case of elastic membranes in Helfrisch's work~\cite{Helfrich1973} where the curvature dependence is expressed as a power-law expansion in the curvature, i.e. the inverse radius, which at second order reads~\cite{Blokhuis1992, Blokhuis1992Rigidity, Aasen2018, Rehner2019}
\begin{equation}\label{eq:sigma_c}
\sigma\left(R_s\right)\simeq\sigma_{0}\left(1-\frac{2\delta}{R_s}+\frac{2\bar{k}+k}{R_s^{2}}\right).
\end{equation}
The flat interface value $\sigma_0$ appears at the leading order, the first order coefficient $\delta$ defines the Tolman length [cf. Fig.~\ref{fig:sketch}$(b)$] and $\bar{k}$ and $k$ are called curvature- and Gaussian-rigidity coefficients, respectively. It has been shown and studied in the literature~\cite{Anisimov2007, Troster2011, Binder2016a} that the Tolman length provides a measure of the ``symmetry'' of the interactions under, e.g., a vapor-liquid exchange transformation. Such symmetry can be readily realized (and broken) in mesoscopic multi-component models~\cite{ShanChen93}. In the next Sections we describe how to tune the degree of asymmetry of the interactions which, in turn, will induce a tuning in the Tolman length and the higher order coefficients.

Several works based on the Density Functional Theory (DFT)~\cite{Boltachev_2003, Blokhuis2013, Wilhelmsen2015, Aasen2018, Rehner2019} have led to expressions for the coefficients $\delta$, $k$ and $\bar{k}$, for realistic multi-phase and multi-component systems. From the numerical perspective, simulations have mostly focused on molecular dynamics (MD) for multi-phase~\cite{Nijmeijer1992, VanGiessen2009, Menzl2016, Langenbach_2018} and multi-component systems~\cite{Yamamoto_2010, Ghoufi_2016}. The Tolman length has been measured in nucleation experiments~\cite{Bruot2016}, and its role was analyzed both in confined geometries~\cite{Kim2018} and in colloidal liquids~\cite{Nguyen2018}. The curvature dependence of the surface tension is paramount in extending Classical Nucleation Theory (CNT). The standard formulation of CNT relies on the \emph{capillary approximation} by which the nucleation free-energy barrier $W$ linearly depends on the flat-interface value of the surface tension $\sigma_0$. Nucleation rates are one of the quantities of interest in CNT with an exponential dependence on $W$, so that, curvature corrections to $\sigma_0$ can quantitatively affect the rate to a large extent. Such considerations are valid for both multi-phase and multi-component systems~\cite{Kalikmanov13} however, much of the literature focuses multi-phase systems for both theory~\cite{Talanquer1995, Tanaka_2015} and experiments~\cite{Bruot2016, Nguyen2018}. Indeed, at present, the application of CNT to multi-component systems is a most challenging yet very important area of research~\cite{Kalikmanov13} in which a sound control over curvature corrections can provide a valuable contribution. {Indeed, in~\cite{Aasen2020} curvature corrections have been used to eliminate a few important inconsistencies of CNT for the case of a propanol-water mixture, e.g. a negative number of molecules in the critical nucleating cluster, highlighting the relevance of the Tolman length and curvature coefficients. Our present mesoscale modelling bears the possibility of extending these results to the hydrodynamic regime where nucleation rates could be consistently predicted also in heterogeneous stress conditions of relevance in several engineering problems.} 

In this work, we study the Tolman length and the higher order corrections using a three-dimensional Shan-Chen multi-component~\cite{ShanChen93, ShanChen94} lattice Boltzmann method (LBM)~\cite{kruger2017lattice, succi2018lattice} by means of an extensive set of hydrostatic simulations. Specifically, we demonstrate that it is possible to tune the value of $\delta$ by ``breaking'', in a controllable way, the symmetry of the system's interactions under the exchange $\AswapB$, thus making a further step with respect to the results already obtained for the multi-phase case~\cite{Lulli_2022} for which the possibility of tuning was left for future works. A strong dependence of the Tolman length on the relative concentrations of the two components has been thoroughly studied in~\cite{Aasen2018} by means of a Square Gradient Theory (SGT) approach which is a first approximation of DFT~\cite{Li_2003}. In the latter case the variation of the curvature corrections are related to the same physical system, while in this work we investigate a parametrization potentially describing different physical systems. We estimate $\delta$ and the combination $2\bar{k} + k$ by leveraging a lattice formulation of the pressure tensor~\cite{Shan08} extended to the multi-component case~\cite{SbragagliaBelardinelli13}. Using the pressure tensor we can compute $\sigma[R]$ (see Fig.~\ref{fig:sketch}$(b)$) following a construction stemming by the mechanic equilibrium condition~\cite{RowlinsonWidom82} which was detailed in~\cite{Lulli_2022}. Most approaches in the field have either leveraged microscopic MD simulations or continuum DFT approaches so that non-equilibrium mescoscopic effects, i.e. hydrodynamics, have mostly been neglected so far. Hence, the present work represents a first step in developing a mesoscale approach for the tuning of the curvature corrections to the surface tension for multi-component systems which can naturally include hydrodynamics. {It would also be interesting to investigate the curvature corrections in different models such as the \emph{color gradient} approach~\cite{Gunstensen_1991,Latva_Kokko_2005,Montessori_2019}, the \emph{free energy} approach~\cite{Swift_1996, Foglino_2017, Tiribocchi_2020} and the \emph{entropic} one~\cite{Mazloomi_M_2015}. Further, it is important to mention that in the case of multi-phase mixtures, in~\cite{Hosseini_2021} results compatible with those in~\cite{Lulli_2022} have been reported, i.e. same power-law behavior of the Tolman length approaching the critical point. Most importantly, the model in~\cite{Hosseini_2021} differs from that in~\cite{Lulli_2022}, for its implementation a for the equation of state, thus providing an important independent validation of the overall approach presented in~\cite{Lulli_2022} and extended here.} {Another interesting perspective is that of studying the relation between curvature corrections and the so-called \emph{near-contact} interactions, e.g. the disjoining pressure that develops when the interfaces of two droplets get close enough and hinders their coalescence. Different lattice Boltzmann models have been devised to correctly capture this kind of interactions~\cite{Benzi_2009, Montessori_2019} and the possibility of an interplay between the Tolman length and the length-scales involved in near-contact interactions will be the focus of future works.}

The paper is organized as follows: we describe in Section~\ref{sec:LBM} the fundamentals of the LBM formulation adopted in this work; in Section~\ref{sec:method} we detail the method used to evaluate the position of the surface of tension $R_s$ and how to ``break'' the symmetry of the interactions in order to achieve a tunable Tolman length; in Section~\ref{sec:res} we report the results followed by the conclusions in Section~\ref{sec:conclusions}. The simulations source code and a Jupyter notebook to reproduce all the results and figures can be found on GitHub \href{https://github.com/lullimat/idea.deploy}{\nolinkurl{https://github.com/lullimat/idea.deploy}}.

\section{Lattice Boltzmann Model}\label{sec:LBM}
The lattice Boltzmann method (LBM) allows to simulate the Navier-Stokes dynamics of a multi-component mixture by means of two coupled forced Boltzmann transport equation acting on a discretized phase-space~\cite{kruger2017lattice, succi2018lattice}. For each component, the single-particle distribution function $f^{\left(\text{J}\right)}(\mathbf{x},\boldsymbol{\xi},t)$ is defined on the the nodes $\{\textbf{x}\}$ of a three-dimensional lattice at discrete times $t$. Hence, one defines the \emph{populations} as the single-particle distribution function evaluated at a given discrete velocity $\boldsymbol{\xi}_i$, i.e. $f_{i}^{\left(\text{J}\right)}(\mathbf{x},t)=f^{\left(\text{J}\right)}(\mathbf{x},\boldsymbol{\xi}_{i},t)$. Remarkably, the convergence to the hydrodynamic limit is very fast even when employing only a few velocity vectors $\{\boldsymbol{\xi}_i\}$ connecting each lattice point to a set neighboring nodes. In this paper we adopt the $D3Q19$ stencil with nineteen descrete velocity vectors $\boldsymbol{\xi}_i$ with $i=0,\ldots,18$. The first two moments of the discretized distribution define the component concentration $n_{\text{J}}=\sum_{i}f_{i}^{\left(\text{J}\right)}$ and the momentum density $n_{\text{J}}\mathbf{u}_{\text{J}}=\sum_{i}f_{i}^{\left(\text{J}\right)}\boldsymbol{\xi}_{i}$, respectively. The lattice transport equation for the J-th component reads
\begin{equation}\label{eq:lbm}
f_{i}^{\left(\text{J}\right)}\left(\mathbf{x}+\boldsymbol{\xi}_{i},t+1\right)-f_{i}^{\left(\text{J}\right)}\left(\mathbf{x},t\right)=\Omega_{i}^{\left(\text{J}\right)}\left(\mathbf{x},t\right)+\left(1-\frac{1}{2\tau_{\text{J}}}\right)F_{i}^{\left(\text{J}\right)}\left(\mathbf{x},t\right)
\end{equation}
where $F_{i}^{\left(\text{J}\right)}\left(\mathbf{x},t\right)$ is the forcing term~\cite{Guo2002} and $\Omega_{i}^{\left(\text{J}\right)}\left(\mathbf{x},t\right)$ is the local collision operator conserving mass and momentum, i.e. $\sum_i \Omega_{i}^{\left(\text{J}\right)} = \sum_i \boldsymbol{\xi}_i \Omega_{i}^{\left(\text{J}\right)} = 0$. Equation~\eqref{eq:lbm} is usually interpreted as implementing two separate steps, namely, i) the \emph{streaming} step represented by the left-hand side by which populations freely stream from one lattice node to the other and ii) the \emph{collision} step represented by the right-hand side which only involves local quantities. The locality of $\Omega_{i}^{\left(\text{J}\right)}$ is one of the main features of LBM which renders the approach particularly amenable to parallel implementations~\cite{kruger2017lattice, succi2018lattice}. More specifically, the right-hand side of~\eqref{eq:lbm} represents is composed by the Bhatnagar-Gross-Krook (BGK)~\cite{Bhatnagar_1954} collision operator
\begin{equation}\label{eq:sm:omega_i}
\Omega_{i}^{\left(\text{J}\right)}\left(\mathbf{x},t\right)=-\frac{1}{\tau_{\text{J}}}\left[f_{i}^{\left(\text{J}\right)}\left(\mathbf{x},t\right)-f_{i}^{\left(\text{eq},\text{J}\right)}\left(n_{\text{J}},\mathbf{u}\right)\right]
\end{equation}
and by the Guo~\cite{Guo2002} forcing term
\begin{equation}
F_{i}^{\left(\text{J}\right)}\left(\mathbf{x},t\right)=w_{i}\left[\frac{1}{c_{s}^{2}}\xi_{i}^{\beta}+\frac{1}{c_{s}^{4}}\left(\xi_{i}^{\alpha}\xi_{i}^{\beta}-c_{s}^{2}\delta^{\alpha\beta}\right)u_{\alpha}\right]F_{\text{J}}^{\beta}\left(\mathbf{x},t\right)
\end{equation}
where repeated Greek indices imply summation. This term is used to implement in the LBM the Shan-Chen~\cite{ShanChen93, ShanChen94} (SC) force $F_{\text{J}}^{\alpha}\left(\mathbf{x},t\right)$ responsible for the formation of stable concentration gradients, i.e. interfaces between the two components. The equilibrium populations $f_{i}^{\left(\text{eq},\text{J}\right)}$ are obtained as a second-order approximation of the Maxwell distribution
\begin{equation}
  f_{i}^{\left(\text{eq},\text{J}\right)}\left(n_{\text{J}},\mathbf{u}\right)=n_{\text{J}}w_{i}\left[1+\frac{1}{c_{s}^{2}}\xi_{i}^{\alpha}u_{\alpha}+\frac{1}{2c_{s}^{4}}\left(\xi_{i}^{\alpha}\xi_{i}^{\beta}-c_{s}^{2}\delta^{\alpha\beta}\right)u_{\alpha}u_{\beta}\right]
\end{equation}
and the equilibrium fluid velocity is computed according to Guo prescription~\cite{Guo2002, kruger2017lattice}
\begin{equation}
  nu^{\mu}=\sum_{\text{J}\in\left\{ \text{A},\text{B}\right\} }\left[\sum_{i=0}^{18}f_{i}^{\left(\text{J}\right)}\xi_{i}^{\mu}+\frac{1}{2}F_{\text{J}}^{\mu}\right].
\end{equation}
Several different approaches for multi-component flows~\cite{kruger2017lattice, succi2018lattice} have been developed for LBM yielding some of the most successful applications of the method. In this paper we show that the SC multi-component model~\cite{ShanChen93,ShanChen94} correctly captures a curvature dependent surface tension while allowing for the tuning of the expansion coefficients, i.e. the Tolman length $\delta$ and the combination of the rigidity constants $2\bar{k} + k$. The main feature of the SC model, allowing for the existence of stable gradients of the concentrations $n_\text{J}(\textbf{x}, t)$, is a force computed on the lattice nodes, which, separating the contribution of each component, reads
\begin{equation}\label{eq:sc_force}
\begin{split}F_{\text{A}}^{\mu}\left(\mathbf{x}\right)= & -Gc_{s}^{2}n_{\text{A}}\left(\mathbf{x}\right)\sum_{a=1}^{18}W\left(|\boldsymbol{\xi}_{a}|^{2}\right)n_{\text{B}}\left(\mathbf{x}+\boldsymbol{\xi}_{a}\right)\xi_{a}^{\mu}\\
 & -G_{\text{AA}}c_{s}^{2}\psi_{\text{A}}\left(\mathbf{x}\right)\sum_{a=1}^{18}W\left(|\boldsymbol{\xi}_{a}|^{2}\right)\psi_{\text{A}}\left(\mathbf{x}+\boldsymbol{\xi}_{a}\right)\xi_{a}^{\mu}\\
F_{\text{B}}^{\mu}\left(\mathbf{x}\right)= & -Gc_{s}^{2}n_{\text{B}}\left(\mathbf{x}\right)\sum_{a=1}^{18}W\left(|\boldsymbol{\xi}_{a}|^{2}\right)n_{\text{A}}\left(\mathbf{x}+\boldsymbol{\xi}_{a}\right)\xi_{a}^{\mu}\\
 & -G_{\text{BB}}c_{s}^{2}\psi_{\text{B}}\left(\mathbf{x}\right)\sum_{a=1}^{18}W\left(|\boldsymbol{\xi}_{a}|^{2}\right)\psi_{\text{B}}\left(\mathbf{x}+\boldsymbol{\xi}_{a}\right)\xi_{a}^{\mu}
\end{split}
\end{equation}
where $\psi_{\text{J}}(\textbf{x}, t) = \psi(n_{\text{J}}(\textbf{x}, t))$ is the so-called pseudopotential, a local function of the concentration $n_{\text{J}}$, implicitly depending on space and time, $c_s = 1/\sqrt{3}$ is the speed of sound, $G$ is the inter-component coupling constant while $G_{\text{AA}}$ and $G_{\text{BB}}$ are the self-coupling constants. If one sets $G = 0$ the two components are completely decoupled and one effectively simulates two parallel multi-phase systems which can display phase separation whenever $G_{\text{AA}},G_{\text{BB}} < G_c$ where $G_c$ is the critical coupling constant whose value depends on the choice of $\psi_{\text{J}}(\textbf{x}, t)$~\cite{ShanChen93,ShanChen94}. A similar approach has been used in~\cite{Benzi_2009, Benzi_2009_1} for the simulation of emulsions and comparison with experimental results~\cite{Derzsi_2017,Derzsi_2018}. The vectors $\boldsymbol{\xi}$ are the discrete forcing directions such that their squared lengths are $|\boldsymbol{\xi}_a|^2 = 1,2$, and $W(1) = 1/6$ and $W(2) = 1/12$ are the weights ensuring 4-th order lattice force isotropy~\cite{Shan06, Sbragaglia07}. The set of the forcing vectors ${\boldsymbol{\xi}_a}$ coincide with that of the lattice velocities ${\boldsymbol{\xi}_i}$ after excluding the ``rest" direction $\boldsymbol{\xi}_0 = (0, 0, 0)$.

Now, a few remarks are in order. From the structure of Eq.~\eqref{eq:sc_force} above it clearly appears that as long as the $G_{\text{AA}} = G_{\text{BB}}$ the system is symmetric, or invariant, under the exchange of the two components $\AswapB$. In order to ``break'' this symmetry one possibility is that of choosing different self-coupling constants $G_{\text{AA}} \neq G_{\text{BB}}$. In order to do this, we adopt the following parametrization
\begin{equation}\label{eq:self_p}
\begin{split}G_{\text{AA}} & =G_{c}\left[1-\delta G_{c}\left(1+\delta G_{\text{AB}}\right)\right]\\
G_{\text{BB}} & =G_{c}\left[1-\delta G_{c}\right]
\end{split}
\end{equation}
where $G_c c_s^2\simeq -2.463$ is the critical value of the self-interaction coupling corresponding to the pseudo-potential $\psi_\text{J} = \exp(-1/n_\text{J})$. It is possible to select other functional forms for $\psi_\text{J}$~\cite{Sbragaglia07, SbragagliaShan10}. The present choice is not meant to fulfill a specific requirement and it is only instrumental for the purpose of analyzing the effects on the Tolman length and the rigidity coefficients of switching from symmetric to asymmetric interactions. By setting $\delta G_c = 0.4$ and $\delta G_\text{AB} > -1$, we assure that $G_{\text{AA}},G_{\text{BB}} > G_c$, i.e. the values of the self-coupling constants are above the critical point so that the gradients in the multi-component system are only due to the inter-component interactions. Indeed, the parameter $\delta G_{\text{AB}}$ estimates the degree of asymmetry of the self interactions, i.e. the ratio $G_{\text{AA}}/G_{\text{BB}}$. One has the following linear relation
\begin{equation}\label{eq:asym_p}
\delta G_{\text{AB}}\left[\frac{\delta G_{c}}{1-\delta G_{c}}\right]=1-\frac{G_{\text{AA}}}{G_{\text{BB}}},
\end{equation}
hence, by setting both positive and negative values we can analyze the behavior of the system around the symmetric case $\delta G_{\text{AB}} = 0$.

The SC force defined in Eq.~\eqref{eq:sc_force} is related to a \emph{lattice} pressure tensor~\cite{Shan08, Belardinelli15, Frometal19, Lulli_2021} that reads
\begin{equation}\label{eq:sm:pt}
  \begin{split}P^{\mu\nu}(\mathbf{x}) & =\left[n_{\text{A}}\left(\mathbf{x}\right)+n_{\text{B}}\left(\mathbf{x}\right)\right]c_{s}^{2}\delta^{\mu\nu}\\
 & +\frac{Gc_{s}^{2}}{2}n_{\text{A}}(\mathbf{x})\sum_{a=1}^{18}W\left(|\boldsymbol{\xi}_{a}|^{2}\right)n_{\text{B}}(\mathbf{x}+\boldsymbol{\xi}_{a})\xi_{a}^{\mu}\xi_{a}^{\nu}\\
 & +\frac{Gc_{s}^{2}}{2}n_{\text{B}}(\mathbf{x})\sum_{a=1}^{18}W\left(|\boldsymbol{\xi}_{a}|^{2}\right)n_{\text{A}}(\mathbf{x}+\boldsymbol{\xi}_{a})\xi_{a}^{\mu}\xi_{a}^{\nu}\\
 & +\frac{G_{\text{AA}}c_{s}^{2}}{2}\psi_{\text{A}}(\mathbf{x})\sum_{a=1}^{18}W\left(|\boldsymbol{\xi}_{a}|^{2}\right)\psi_{\text{A}}(\mathbf{x}+\boldsymbol{\xi}_{a})\xi_{a}^{\mu}\xi_{a}^{\nu}\\
 & +\frac{G_{\text{BB}}c_{s}^{2}}{2}\psi_{\text{B}}(\mathbf{x})\sum_{a=1}^{18}W\left(|\boldsymbol{\xi}_{a}|^{2}\right)\psi_{\text{B}}(\mathbf{x}+\boldsymbol{\xi}_{a})\xi_{a}^{\mu}\xi_{a}^{\nu}.
\end{split}
\end{equation}
We wish to highlight that the tensor in the Eq.~\eqref{eq:sm:pt} is such that the flat-interface mechanical equilibrium condition, i.e. constant normal component $P_{\text{N}}(x) = p_0$ throughout the interface, is obeyed \emph{on the lattice} with a value of $p_0$ that is constant to machine precision. This property has allowed for an extremely precise estimation of the coexistence curve in the multi-phase case~\cite{Shan08, SbragagliaShan10,Frometal19} and it is one of the building blocks for the results presented in this paper. By performing the Taylor expansion of Eq.~\eqref{eq:sm:pt} one obtains, at the leading order, the bulk pressure
\begin{equation}\label{eq:bulkp}
  \begin{split}P(\mathbf{x}) & =\left[n_{\text{A}}\left(\mathbf{x}\right)+n_{\text{B}}\left(\mathbf{x}\right)\right]c_{s}^{2}+Gc_{s}^{2}e_{2}n_{\text{A}}(\mathbf{x})n_{\text{B}}(\mathbf{x})\\
 & +\frac{G_{\text{AA}}c_{s}^{2}e_{2}}{2}\psi_{\text{A}}^{2}(\mathbf{x})+\frac{G_{\text{BB}}c_{s}^{2}e_{2}}{2}\psi_{\text{B}}^{2}(\mathbf{x}),
\end{split}
\end{equation}
where $e_2 = \sum_{\textbf{e}_a} W(|\textbf{e}_a|^2)e_a^x e_a^x=1$~\cite{Shan06, SbragagliaBelardinelli13, Lulli_2021} for the values of the weights used in this work. The first line represents the ideal gas contribution plus the inter-component interaction contribution while the second line yields the sum of the self-interactions ones. Considering the combination of the ideal and self-interaction parts each component can independently display phase separation whenever $G_\text{AA},G_\text{BB} < G_c$~\cite{ShanChen93}.

The SC model has been widely used to model complex fluids with a non-trivial impact on the study of the interface physics, one may cite heterogeneous cavitation~\cite{Falcucci13a} and emulsion rheology physics~\cite{Lulli_2018}, also in presence of complex boundary conditions~\cite{Derzsi_2018}. The ability to model and tune the Tolman length and the rigidity coefficients in LBM allows to effectively tackle the study of nucleation and cavitation phenomena in the mesoscale regime for multi-component systems, while providing a computationally efficient tool allowing for a direct bridge to experiments.

\section{Method}\label{sec:method}
In Section~\ref{sec:Intro} we briefly discussed that in a multi-component system the free energy needs to be independent on the choice of the position $R$ of an arbitrary dividing spherical surface. Such a condition yields the generalized Laplace law~\cite{GibbsCollected1948, Buff1951, RowlinsonWidom82, Rowlinson1984}
\begin{equation}\label{eq:LaplaceGen}
\Delta P=\frac{2\sigma\left[R\right]}{R}+\left[\frac{\text{d}\sigma}{\text{d}R}\right]
\end{equation}
with $\sigma[R]$ the generalized surface tension and its {notional derivative~\cite{RowlinsonWidom82}} $[\text{d}\sigma/\text{d}R] = \sigma'[R]$ and $\Delta P = P_{\text{in}} - P_{\text{out}}$, with $P_{\text{in}}$ and $P_{\text{out}}$ the values of the bulk pressure in the center of the droplet and far away from the interface, respectively (see Fig.~\ref{fig:sketch}$(a)$). The function $\sigma[R]$ is convex and at its minimum Eq.~\eqref{eq:LaplaceGen} reduces to the usual Laplace law. The condition $\sigma'[R]|_{R=R_s} = 0$ defines the position of the surface of tension $R_s$. Hence, comparing Eqs.~\eqref{eq:sigma_c} and~\eqref{eq:LaplaceGen}, it follows that at second order in $R_s^{-1}$ the latter reads \begin{equation}\label{eq:Laplace2nd}
    \Delta P = \frac{2\sigma_s(R_s)}{R_s}\simeq \frac{2\sigma_0}{R_s}\left(1 - \frac{2\delta}{R_s} + \frac{2\bar{k}+k}{R_s^{2}}\right).
\end{equation}
In order to estimate the Tolman length we simulate droplets with $n_\text{A}$- and $n_\text{B}$-rich bulks for different values of the asymmetry parameter $\delta G_{\text{AB}}$. Further, we compute the deviations from the Laplace law using the surface of tension radius $R_s$ which is used to determine the droplets sizes. In order to estimate $R_s$ from the simulations we use a construction presented in~\cite{RowlinsonWidom82} which only employs the mechanic equilibrium condition $\partial_\mu P^{\mu\nu} = 0$. The same arguments have been adopted in the case of the multi-phase SC model in~\cite{Lulli_2022} where they are described in details. Here, we limit our discussion to the most important steps. Let us consider the following decomposition of the pressure tensor 
\begin{equation}\label{eq:pt_projection}
    P^{\mu\nu} = P_{\text{N}}\delta^{\mu\nu} - (P_{\text{N}} - P_{\text{T}})q^{\mu\nu},
\end{equation}
where $P_{\text{N}}$ and $P_{\text{T}}$ are the (locally) normal and tangential components to the droplet interface, respectively. The projector along the tangential direction is defined as $q^{\mu\nu}=\delta^{\mu\nu} - n^\mu n^\nu$ where $n^\mu$ is the normal vector to the interface which is given by the direction of the largest gradient.
The mechanic equilibrium condition reads
\begin{equation}
    \partial_{\mu}P^{\mu\nu}= n^{\nu}n^{\mu}\partial_{\mu}P_{\text{N}}+n^{\nu}\partial_{\mu}n^{\mu}\left(P_{\text{N}}-P_{\text{T}}\right) = 0.
\end{equation}
In three dimensions one has $n^{\nu}\partial_{\mu}n^{\mu}=2n^{\nu}/{r}$, where $r$ is the value of the radial coordinate. Selecting the normal/radial direction to be parallel to the $x$-axis, i.e. $n^\mu = e_x^\mu$ yields
\begin{equation}\label{eq:mech_eq_radial}
    \frac{\text{d}}{\text{d}r}P_{\text{N}}\left(r\right)+\frac{2}{r}\left[P_{\text{N}}\left(r\right)-P_{\text{T}}\left(r\right)\right] =0.
\end{equation}
Upon multiplication by $r^n$ followed by some derivatives rearrangements it possible to obtain a sequence of identities
\begin{equation}\label{eq:n_mech_eq}
    \frac{\text{d}}{\text{d}r}[r^{n}P_{\text{N}}(r)]=r^{n-1}[(n-2)P_{\text{N}}(r)+2P_{\text{T}}(r)].
\end{equation}
Finally, after introducing the pressure-jump function $P_{\text{J}}(r;R)=P_{\text{in}}-(P_{\text{in}}-P_{\text{out}})\theta(r-R)$, where $\theta(r - R)$ is the Heaviside function, one can subtract the integral between $R_{\text{in}}$ and $R_{\text{out}}$ of Eq.~\eqref{eq:n_mech_eq} and that of $r^n\,P_{\text{J}}(r;R)$. After setting $n = 2$ one obtains the following expression for the pressure jump $\Delta P=P_{\text{in}}-P_{\text{out}}$ across the interface
\begin{equation}
\begin{split}\Delta P & =\frac{2}{R^{2}}\int_{R_{\text{in}}}^{R_{\text{out}}}\text{d}r\;r[P_{\text{J}}(r;R)-P_{\text{T}}(r)]\\
 & =\frac{2\sigma[R]}{R}+\left[\frac{\text{d}\sigma}{\text{d}R}\right].
\end{split}
\end{equation}
It is possible to extract the expressions for $\sigma[R]$ and $[\mbox{d}\sigma/\mbox{d}R]$~\cite{RowlinsonWidom82} obtaining
\begin{equation}\label{eq:gen_sigma}
    \sigma[R]=\int_{0}^{+\infty}\text{d}r\;\left(\frac{r}{R}\right)^{2}[P_{\text{J}}(r;R)-P_{\text{T}}(r)],
\end{equation}
\begin{equation}\label{eq:gen_d_sigma}
    \left[\frac{\text{d\ensuremath{\sigma}}}{\text{d}R}\right]=-\frac{2}{R^{3}}\int_{0}^{+\infty}\text{d}r\;r(r-R)[P_{\text{J}}\left(r;R\right)-P_{\text{T}}\left(r\right)],
\end{equation}
where we also considered the limits $R_{\text{in}} \to0$ and $R_{\text{out}}\to\infty$. In order to estimate the position of the surface of tension $R_s$, we interpolate the position of the minimum of Eq.~\eqref{eq:gen_sigma} after evaluating the expression by means of the SC lattice pressure tensor in Eq.~\eqref{eq:sm:pt}, integrating along the $x$ axis so that $P_{\text{N}} = P^{xx}$ and $P_{\text{T}} = P^{yy} = P^{zz}$. The pressure jump across the droplets interfaces $\Delta P$ and the position of the surface of tension $R_s$ are the key quantities in our analysis allowing us, by means of hydrostatic simulations of droplets of different sizes, to estimate the curvature dependence of the surface tension as $\sigma(R_s)/\sigma_0 = R_s\,\Delta P / 2\sigma_0$. We remark that other choices are possible for the dividing surface, such as the total equimolar interface~\cite{Aasen2018}, however, such choices allow the {notional derivative~\cite{RowlinsonWidom82}} in Eq.~\eqref{eq:LaplaceGen} to play a non-trivial role in the estimation of the coefficients, whereas the surface of tension allows to directly estimate the function $\sigma(R_s)$. Moreover, more than one definition for an equimolar radius is possible, appearing as a further dependence for the rigidity coefficients and not for the Tolman length~\cite{Boltachev_2003, Aasen2018}. All in all, using the surface of tension, as already done in~\cite{Yamamoto_2010}, allows for a simpler analysis of the surface tension curvature dependence.

\section{Results}\label{sec:res}
\begin{figure}[!t]
    \centering
    \includegraphics[scale = 0.45]{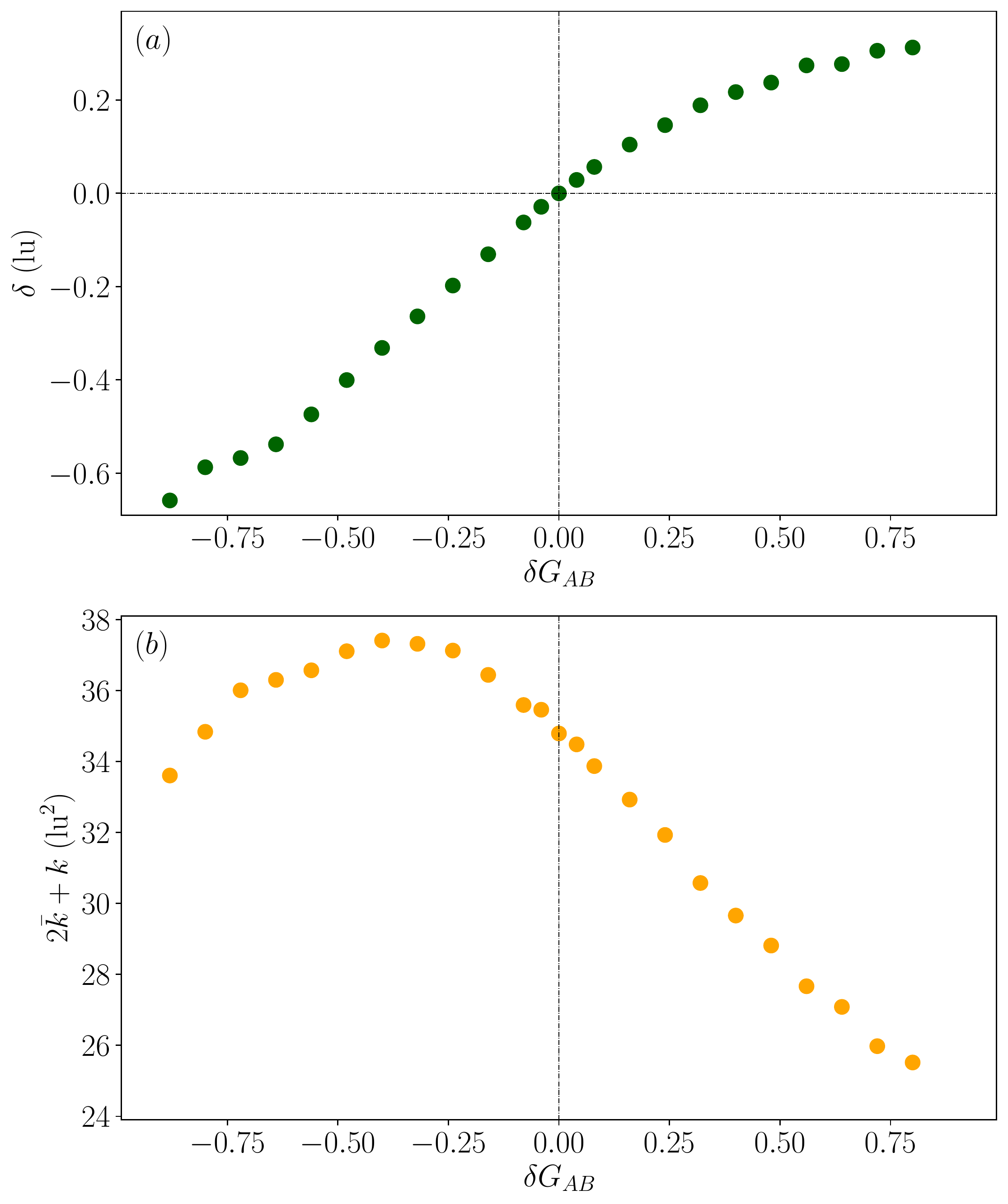}
    \caption{$(a)$: Values of the Tolman length $\delta$ as a function of the asymmetry parameter $\delta G_\text{AB}$, vanishing in the symmetric case $\delta G_\text{AB} = 0$. $(b)$: Values of the combination of the rigidity coefficients $2\bar{k} + k$ as a function of $\delta G_\text{AB}$ displaying a non-monotonic behavior. {$\delta$, and the curvature coefficients are reported in lattice units, i.e. lu and lu$^2$.}}\label{fig:curv_coeffs}
\end{figure}

The simulations source code can be found on GitHub \href{https://github.com/lullimat/idea.deploy}{\nolinkurl{https://github.com/lullimat/idea.deploy}}~\cite{sympy, scipy, numpy0, numpy1, scikit-learn, matplotlib, ipython, pycuda_opencl}. A Jupyter notebook~\cite{ipython} is available from the ``idea.deploy" framework to reproduce the results and the plots reported in this paper. {The code provided for the multi-component model relies on a straight-forward implementation, i.e. not highly optimized, with a set of kernels that can be compiled either in CUDA or OpenCL. This version still does not leverage the automatic code generation already implemented for the multi-phase case which will soon be extended to the multi-component one. In order to give an estimate of the needed simulation time, one needs roughly 5hrs on a Tesla P100 or 1.5hrs on a Tesla A100, for executing the simulations in order to obtain the data for one of the points in Fig.~\ref{fig:curv_coeffs} with $L\leq 213$lu for a maximum RAM usage of 8GB per simulation. This is not the full size range presented here which requires 16GB of RAM per simulation. Hence, with the former constraint one would need roughly 5 days on one P100 and 1.5 days on a A100. The largest system size is the most challenging, not only because it requires more resources, but also because the actual convergence is slower.} The strategy used for the simulations closely follows our previous contribution~\cite{Lulli_2022}. Here we report some details for completeness. We simulate three-dimensional droplets in a cubic system of linear size $L$ with periodic boundary conditions using the D3Q19 discrete velocity set with $c_s^2=1/3$~\cite{kruger2017lattice, succi2018lattice}. We adopt $\psi_\text{J} = \exp(-1/n_\text{J})$~\cite{ShanChen94} as the pseudo-potential function for the self-interaction part in Eq.~\eqref{eq:sc_force}. Other definitions of $\psi_\text{J}$ have been used in the literature, however, the present choice is just as suitable for our primary objective, i.e. a first exploration of the tunability of the curvature corrections coefficients. The asymmetry parameter varies in the range $\delta G_{\text{AB}} \in \{0, \pm 0.04, \pm 0.08, \pm 0.16, \pm 0.24, \pm 0.32, \pm 0.40, \pm 0.48, \pm 0.56, \pm 0.64, \pm 0.80, \pm 0.88\},\;$ $\delta G_c$ is set to 0.4, $G_c c_s^2 \simeq -2.463$ and the inter-component coupling is set to $Gc_s^2 = 0.5$ (cf. Eq.~\eqref{eq:sc_force}). The value of $L$ is chosen to be an odd number so that the center of mass of the system exactly falls on the coordinates of a node. The simulated system sizes are $L\in \{25, 31, 37, 41, 51, 57, 65, 97, 127, 151, 213, 301\}${lu, where ``lu'' stands for lattice units}. The radial concentration fields $n_{\text{J}}(r)$ are initialized to the following profile
\begin{equation}
n_{\text{J}}(r,R)=\frac{1}{2}(n_{\text{J,in}}+n_{\text{J,out}})-\frac{1}{2}(n_{\text{J,in}}-n_{\text{J,out}})\tanh(r-R),
\end{equation}
where the inner $n_{\text{J,in}}$ and outer $n_{\text{J,out}}$ densities are set to the steady-state values obtained from the simulations of a flat interface system and the initial value of the radius is set to maintain a fixed aspect ratio $R = L/4$ for all simulations. The radial coordinate $r$ is computed taking the center of the system as the origin. The values $P_{\text{in}}$ and $P_{\text{out}}$ are evaluated in the middle of the system $(\lfloor L/2 \rfloor, \lfloor L/2 \rfloor, \lfloor L/2 \rfloor)$ and at the farthest corner $(L - 1, L - 1, L - 1)$, respectively. The outcome of the simulations is analyzed only if all the coordinates of the center of mass lie within a distance of $10^{-3}$ from the center of the domain. We use two convergence criteria for the simulations, both comparing quantities at a time distance $\delta t = 2^{11}$: i) we consider the relative variation of the $\Delta P$ with respect to the previous configuration, and when the latter is such that $|\Delta P(t) - \Delta P(t + \delta t)|/\Delta P(t) < 10^{-5}$ the simulation is considered as converged; ii) we consider the magnitude $\delta u$ of the spatial average of the difference between the components of two velocity fields, $\delta u=L^{-3}\sum_{\textbf{x}}\sum_{\alpha}|u^{\alpha}(\textbf{x},t+\delta t)-u^{\alpha}(\textbf{x},t)|$ so that the simulation is considered as converged when $\delta u< 10^{-12}$. Meeting only one of the two criteria is enough to finalize the simulation.

The set of simulations for the flat interface has been performed on a three-dimensional domain of sizes $L_x = 127, L_y = L_z = 5$ and the concentration profiles are initialized according to
\begin{equation}
\begin{split}n_{\text{J}}\left(x,x_{0},w\right)= & \frac{1}{2}\left(n_{\text{J,h}}+n_{\text{J,l}}\right)\\
- & \frac{1}{2}\left(n_{\text{J,h}}-n_{\text{J,l}}\right)\tanh\left[x-\left(x_{0}-\frac{w}{2}\right)\right]\\
+ & \frac{1}{2}\left(n_{\text{J,h}}-n_{\text{J,l}}\right)\left\{ \tanh\left[x-\left(x_{0}+\frac{w}{2}\right)\right]+1\right\} ,
\end{split}
\end{equation}
where $x_0$ is the center of the strip and $w = L_x/2$ its width. As a first approximation, in the presence of self-interactions, the flat interface concentrations, $n_{\text{J,h}}$ and $n_{\text{J,l}}$ for the \emph{high} and \emph{low} value respectively, can be computed using the purely repulsive result of Eq.(43) in~\cite{Benzi_2009} with the substitution $\theta(\tau) \to 1$ to take into account Guo's forcing~\cite{Guo2002}. Such values are then used to initialize the flat interface profile which is simulated until the steady state is reached. The final concentrations are then used to initialize the spherical interface simulations. This procedure proves to be effective in providing a good starting point for the droplets simulations which are able to reach the steady state in a reasonable time avoiding large pressure waves originating by a less precise estimation of the initial concentration values.
\begin{figure}[!t]
    \centering
    \includegraphics[scale = 0.45]{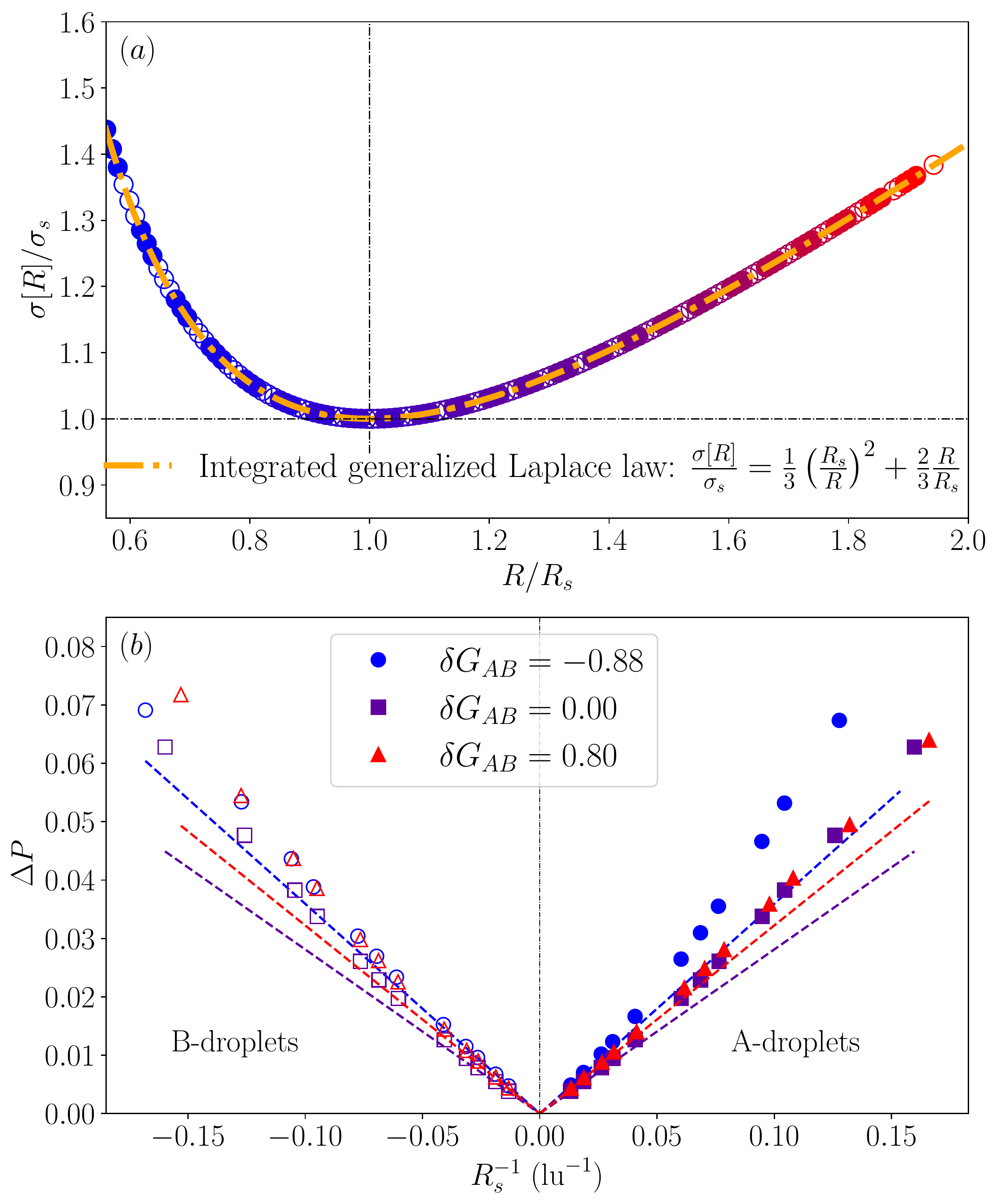}
    \caption{$(a)$: Collapse of the rescaled generalized surface tension $\sigma[R]/R_s$ as a function of the rescaled arbitrary dividing surface position $R/R_s$ for all the simulated $n_\text{A}$- and $n_\text{B}$-rich droplets, related to full and empty circles respectively, for all the values of $\delta G_\text{AB}$, represented by different colors, and all system sizes $L$, displaying a good superposition to the expected ``universal'' behavior. $(b)$: {Values of the pressure difference $\Delta P$ as a function of the curvature (negative for B-rich droplets) at the surface of tension $R_s^{-1}$} for the extrema of the asymmetry parameter $\delta G_\text{AB} = \pm 0.88$ and the fully symmetric case $\delta G_\text{AB} = 0$. Corrections to the {expected Laplace law (dashed lines)} are visible and asymmetric when changing the sign of $\delta G_\text{AB}$. {The curvature $R_s^{-1}$ is reported in inverse lattice units, i.e. lu$^{-1}$}}\label{fig:gen_sigma_laplace}
\end{figure}
\begin{figure}[!t]
    \centering
    \includegraphics[scale = 0.45]{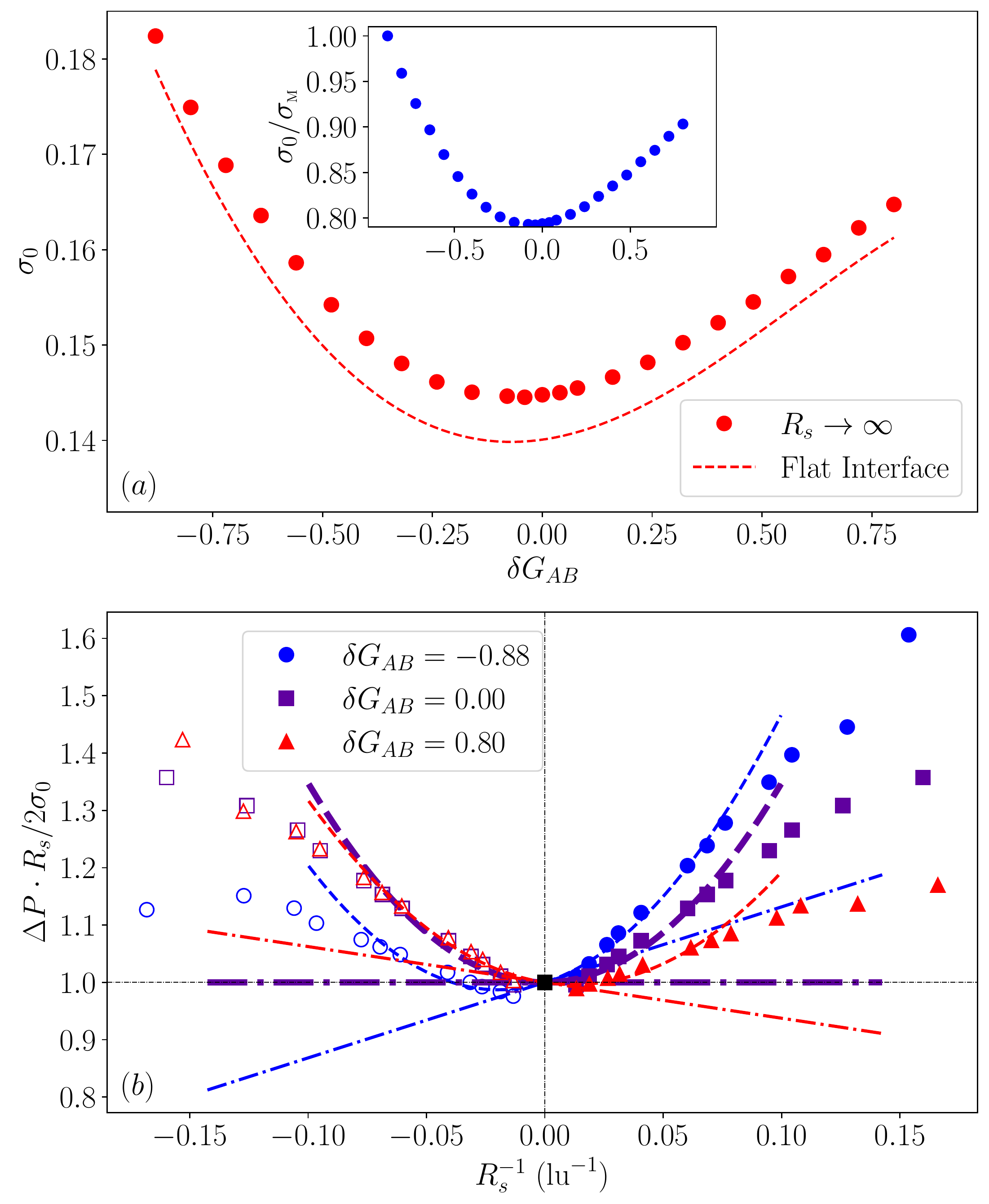}
    \caption{$(a)$: Values of the surface tension as estimated from the droplets simulations in the flat interface limit (symbols) compared to the values computed from flat interface simulations. The discrepancy is at most $3\times 10^{-2}$. The inset displays the variation relative to the maximum which does not exceed $2\times 10^{-1}$. $(b)$: Curvature dependence of the surface tension $\sigma(R_s)/\sigma_0 = \Delta P \cdot R_s / 2\sigma_0$ for three different choices of the asymmetry parameter $\delta G_\text{AB}$. Dashed-dotted straight lines indicate the results for the linear term while the dashed parabolas indicate the second order approximation. Thicker lines are used for the symmetric case $\delta G_\text{AB} = 0$. {The curvature $R_s^{-1}$ is reported in inverse lattice units, i.e. lu$^{-1}$}}\label{fig:sigma0_laplace}
\end{figure}

As a first result we report in Fig.~\ref{fig:gen_sigma_laplace}$(a)$ the data for the rescaled generalized surface tension $\sigma[R]/\sigma_s$, where $\sigma_s = \sigma[R_s] = \sigma(R_s)$ is the value at the minimum, as a function of the normalized position of the arbitrary diving surface $R/R_s$. It is possible to compare these data with an analytical expression obtained from the integration of the generalized Laplace law: we can rewrite Eq.~\eqref{eq:LaplaceGen} as $R^{2}\Delta P=\text{d}\left[R^{2}\sigma\left[R\right]\right]/\text{d}R$ and integrate from $R_s$ to $R$ and obtain~\cite{RowlinsonWidom82} the expression $\frac{\sigma\left[R\right]}{\sigma_{s}}=\frac{1}{3}\left(\frac{R_{s}}{R}\right)^{2}+\frac{2}{3}\frac{R}{R_s}$. The latter one is referred to as ``universal'' in~\cite{Troster2011}, i.e. not depending on temperature or on the droplet size, mirroring that $\sigma[R]$ depends on the arbitrary value of $R$. In Fig.~\ref{fig:gen_sigma_laplace}$(a)$ we compare the results obtained from the entire set of simulations with the analytical prediction, yielding a good agreement. This result allows us to determine the positions of the surface of tension $R_s$ from the minima of the generalized surface tension curves. Fig.~\ref{fig:gen_sigma_laplace}$(b)$ displays the points $(R_s^{-1}, \Delta P)$, i.e. the Laplace law, for the symmetric interactions with $\delta G_\text{AB} = 0$, and the two most asymmetric cases $\delta G_\text{AB} = \pm 0.88$. All curves converge to the slope expected from the flat interface surface tension, i.e. $2\sigma_0$, while sizeable corrections are visible for smaller droplets.

We analyze in further details the surface tension $\sigma_0$ as computed from (i) flat interface simulations and (ii) from the $R_s \to \infty$ limit of the data obtained from the droplets simulations. Fig.~\ref{fig:sigma0_laplace}$(a)$ displays the results for different values of $\delta G_\text{AB}$ reaching a minimum near the symmetric case $\delta G_\text{AB} = 0$ and increasing at the boundary of the interval $\delta G_\text{AB} = \pm 0.88$. We report using circles the $R_s\to \infty$ data while we those for the flat interface simulations are reported in dashed. In the first case we use parabolic fits of the quantity $\sigma(R_s) = \Delta P \cdot R_s / 2$ to estimate the value of $\sigma_0$ in the $R_s\to \infty$ limit, while for the flat interface we use the mechanical definition of the surface tension
\begin{equation}
    \sigma_0 = \int_{L_x/2}^{L_x} \mbox{d}x [P_{\text{N}}(x) - P_{\text{T}}(x)],
\end{equation}
where $P_{\text{N}}(x) = P^{xx}(x)$ and $P_{\text{T}}(x) = P^{yy}(x)$ have been obtained from the lattice pressure tensor~\eqref{eq:sm:pt}. The relative difference between the two estimates for $\sigma_0$ never exceeds $3.7\times 10^{-2}$, while the relative difference between the minimum and the maximum values, $\sigma_\text{m}$ and $\sigma_\text{M}$ respectively, as a function of $\delta G_\text{AB}$ is bounded at $1 - \sigma_\text{m}/\sigma_\text{M}\sim 2\times 10^{-1}$. We continue with the analysis of the curvature corrections to the flat interface value of the surface tension $\sigma_0$. In Fig.~\ref{fig:sigma0_laplace}$(b)$ we report the data for $\sigma(R_s)/\sigma_0$ as estimated from the droplet simulations through the ratio $\Delta P \cdot R_s/2\sigma_0$ (cf. Eq.~\eqref{eq:Laplace2nd}). We wish to remark that the values of $\sigma_s = \sigma(R_s)$ estimated from the pressure jump $\Delta P$ and those obtained from the minimum of $\sigma[R]$ in Eq.~\eqref{eq:gen_sigma} have a relative difference of at most $4\times 10^{-3}$. We only report the symmetric, $\delta G_\text{AB} = 0$, as well as the most asymmetric cases, $\delta G_\text{AB}=\pm 0.88$ for ease of reading. We perform quadratic fits in order to estimate the first and second order coefficients. After normalizing the coefficients by the zero-th order one, i.e. $\sigma_0$, one obtains $-2\delta$ and $2\bar{k} + k$ respectively. The values of $\sigma_0$ are reported in Fig.~\ref{fig:sigma0_laplace}. Dash-dotted lines represent the result for the linear coefficient. The thicker line indicates the symmetric case which has a negligible slope, i.e. vanishing Tolman length. On the other hand, the two asymmetric cases display finite slopes of opposite signs, signaling a change in the of the Tolman length. We report in dashed lines the results for the fits of the full parabola which give a good approximation in the $R_s^{-1}\to 0$ limit.

Finally, Fig.~\ref{fig:curv_coeffs}$(a)$ and $(b)$ display the Tolman length $\delta$ and the rigidity coefficients combination $2\bar{k} + k$, respectively, as a function of the asymmetry parameter $\delta G_\text{AB}$. As already visible from Fig.~\ref{fig:sigma0_laplace} the sign of the Tolman length changes with the sign of $\delta G_\text{AB}$ with an almost monotonic dependence. {Moreover, one can notice that the the absolute value of $\delta$ is larger for negative $\delta G_\text{AB}$. This can be understood given that for $\delta G_\text{AB}<0$ the A-component gets closer to the critical point of the self interactions. While these are still too weak to induce phase separation, they exert a stronger effect on the interface with respect to $\delta G_\text{AB}>0$ branch for which the interface features are set, almost entirely, by the inter-component interactions. A non-trivial competition between inter-component and self interactions for the formation of the interface is probably responsible for both the presence of a maximum for $2\bar{k} + k$ and the non-monotonic behavior of the first derivative clearly visible around $\delta G_{\text{AB}} \simeq -0.75$. A theoretical prediction for both the Tolman length and the higher order curvature coefficients will be paramount to fully capture this competition among interactions.} On the other hand, the rigidity coefficients display a non-monotonic behavior, reaching a maximum for $\delta G_\text{AB}\simeq -0.40$. We notice that the results do not show symmetry under sign exchange for $\delta G_\text{AB}$ mirroring the asymmetric change in the interactions under exchange of the two components $\AswapB$. The relative change of $\delta$ is around $1 - \delta_\text{m}/\delta_\text{M} \sim 1.5$, with $\delta_\text{m}$ and $\delta_\text{M}$ the norm of the minimum and the maximum values, respectively. Hence $1 - \delta_\text{m}/\delta_\text{M}$ is far larger than $1 - \sigma_\text{m}/\sigma_\text{M}$, so that one variation is weakly dependent on the other. {A few remarks on the dependence of the results on the simulations parameters are in order. Let us begin from the system size $L$ dependence: limiting the set of simulations to a value $L\leq 213$lu still yields a consistent curve for $\delta$ while the estimates for $2\bar{k}+k$ change by roughly 20\%. This is due to the fact that the points closest to the flat interface limit $R_s^{-1}\simeq 0$ are the most significant for getting a reliable estimate of the second order coefficient for the curvature corrections. Furthermore, the value of the inter-component coupling $G$ has been chosen so that the largest spurious currents is of order $O(10^{-3}\mbox{Ma})$, where Ma is the Mach number, which is weak enough not to affect the estimations of $\Delta P$. Moreover, since the magnitude of the spurious currents is correlated to the surface tension~\cite{Sbragaglia07}, the relatively small variation of $\sigma_0$ reported in Fig.~\ref{fig:sigma0_laplace}$(a)$ assures that the spurious currents consistently stay at the same order in the whole range of $\delta G_{AB}$. Finally, in a three-dimensional system, the initial droplet size ratio $L/R=4$ is large enough to avoid the spurious currents effect to propagate through the periodic boundaries.} Future works will aim at completely disentangle the variations of the two quantities making them independent.

\section{Conclusions}\label{sec:conclusions}
In the present work we demonstrate, by means of an extensive set of simulations of a two-component system, that (i) the Shan-Chen~\cite{ShanChen93, ShanChen94} multi-component model is able to capture the curvature corrections to the surface tension and (ii) it naturally allows for a straightforward method for tuning both the Tolman length and the rigidity coefficients in a wide range of values, while keeping the surface tension in a relatively narrow range. Specifically, this is obtained by tuning the degree of asymmetry~\cite{Anisimov2007} of each component self-interaction~\cite{Benzi_2009} while keeping the cross-component interaction constant. By this method we demonstrate how the Tolman length, i.e. the first-order curvature correction of the surface tension in the flat interface limit, can be made to vanish in a continuous way by restoring the symmetry of the interaction under exchange of the two components $\AswapB$. The tuning of the Tolman length, especially by means of the tuning of the relative concentrations of the two components has been thoroughly studied in the context of Density Functional Theory approaches~\cite{Aasen2018}. While those studies represent a variation of the curvature corrections for the same physical system, here we chose, as a first instance, a parametrization potentially describing different physical systems.

This represents a first step for the tuning of the curvature corrections in order to correctly model different realistic systems. Further studies will address the same results seeking an analytical control also for the multi-phase systems for which recent results~\cite{Lulli_2022} have already demonstrated the existence of the Tolman length and its temperature dependence in the Shan-Chen multi-phase model. This research direction holds the potential to allow a more straightforward approach for the study and modelling of nucleation and cavitation problems taking naturally into account the hydrodynamic contributions, while offering, at the same time, a very computationally efficient method capable of dealing with complex and realistic boundary conditions. The simulations source code and a Jupyter notebook to reproduce all the results and figures can be found on GitHub \href{https://github.com/lullimat/idea.deploy}{\nolinkurl{https://github.com/lullimat/idea.deploy}}.

\begin{acknowledgement}
The authors wish to thank {\O}ivind Wilhelmsen for useful discussion. Luca Biferale thankfully acknowledges the hospitality from the Department of Mechanics and Aerospace Engineering of Southern University of Science and Technology. This work was supported by National Science Foundation of China
 Grants~12050410244, 91741101 and 91752204, by Department of Science and Technology of
 Guangdong Province Grant No.~2019B21203001, Science and Technology
 Innovation Committee of Shenzhen Grant No.~K19325001, and from the European Research Council (ERC) under the European Union’s Horizon 2020 research and innovation programme (grant agreement No 882340).
\end{acknowledgement}

{\section*{List of Symbols}\label{sec:syms}
Symbols are reported according to their order of appearance. Bold font symbols refer to tensorial quantities that can be indexed through Greek letters.
\newline
\newline
\begin{tabular}{p{0.175\textwidth}p{0.825\textwidth}}
    $n_\text{J}$ & \quad concentration of the J-th fluid component \\
    $\mathbf{u}$ & \quad fluid velocity field \\
    $n_\text{J,s}$ & \quad saturation concentration of the J-th fluid component \\
    $\sigma$ & \quad surface tension \\
    $n_\text{J,b}$ & \quad concentration of the J-th fluid component in the bulk of a J-rich droplet\\
    $n_\text{J,out}$ & \quad concentration of the J-th fluid component outside of a J-rich droplet\\
    $z_s$ & \quad surface of tension position for a flat interface\\
    $R$ & \quad position of an arbitrary dividing surface for a droplet\\
    $\sigma[R]$ & \quad generalized surface tension\\
    $\left[\frac{\text{d}\sigma}{\text{d}R}\right]$ & \quad notional derivative of the generalized surface tension\\
    $\{A,B\}$ & \quad components labels for a binary mixture\\
    $R_s$ & \quad position of the surface of tension for a droplet\\
    $\Delta P$ & \quad pressure jump between the inside and outside of a droplet\\
    $\sigma(R)$ & \quad curvature dependent surface tension\\
    $\sigma_0$ & \quad flat interface surface tension\\
    $\delta$ & \quad Tolman length\\
    $\bar{k}$,$k$ & \quad curvature- and Gaussian-rigidity coefficients\\
    $f^{\left(\text{J}\right)}(\mathbf{x},\boldsymbol{\xi},t)$ & \quad single-particle distribution function for the J-th component\\
    $\{\mathbf{x}\}$ & \quad set of discrete lattice points\\
    $\boldsymbol{\xi}$ & \quad particle peculiar velocity\\
    $t$ & \quad time\\
    $\{\boldsymbol{\xi}_i\}$ & \quad discrete velocity set or stencil\\
    $f^{\left(\text{J}\right)}_i$ & \quad $i$-th population for the J-th component\\
    $n_\text{J} \mathbf{u}_\text{J}$ & \quad J-th component momentum\\
    $F^{(\text{J})}_i$ & \quad forcing term in the Lattice Boltzmann equation for the $i$-th population of the J-th component\\
    $\Omega_i^{(\text{J})}$ & \quad collision operator for the $i$-th population of the J-th component\\
    $\tau_{\text{J}}$ & \quad BGK relaxation for the J-th component\\
    $f_{i}^{\left(\text{eq},\text{J}\right)}$ & \quad discrete equilibrium distribution for the $i$-th population of the J-th component\\
    $c_s^2$ & \quad square of the speed of sound\\
    $w_i$ & \quad weight associated to the discrete velocity $\boldsymbol{\xi}_i$\\
    $\mathbf{F}_\text{J}$ & \quad force exerted on the J-th component\\
    $G$ & \quad inter-component coupling constant\\
    $G_\text{AA}$, $G_\text{BB}$ & \quad components self-coupling constants\\
    $\psi_\text{J}$ & \quad pseudo-potential function for self interactions\\
    $W$ & \quad weights associated to the forcing directions\\
\end{tabular}
\newline
\phantom{~}\noindent
\begin{tabular}{p{0.175\textwidth}p{0.825\textwidth}}
    $G_c$ & \quad critical value for phase separation for the self interactions for the given $\psi_\text{J}$\\
    $\delta G_c$ & \quad relative distance from the critical point\\
    $\delta G_\text{AB}$ & \quad asymmetry parameter estimating the relative difference of $G_\text{AA}$, $G_\text{BB}$\\
    $P^{\mu\nu}$ & \quad (lattice) pressure tensor\\
    $P_\text{N}$ & \quad normal component of the pressure tensor to an interface\\
    $p_0$ & \quad bulk pressure for a flat interface\\
    $P$ & \quad bulk pressure\\
    $e_2$ & \quad second order isotropy coefficient\\
    $\mathbf{q}$ & \quad projector onto the tangetial direction to an interface\\
    $\mathbf{n}$ & \quad normal to an interface\\
    $r$ & \quad radial coordinate\\
    $P_\text{in}$, $P_\text{out}$ & \quad bulk pressure inside and outside of a droplet\\
    $P_\text{J}$ & \quad pressure jump function \\
    $\theta$ & \quad Heaviside step function\\
    $L$, $L_x$, $L_y$, $L_z$ & \quad linear system size according to the direction\\
    $n_\text{J,in}$, $n_\text{J,out}$ & \quad initial concentrations inside and outside a droplet for the J-th component\\
    $n_\text{J,h}$, $n_\text{J,l}$ & \quad high and low initial concentrations values for the J-th component with a flat interface\\
    $x_0$, $w$ & \quad center and width of the initial profile for the flat interface\\
    $\sigma_\text{m}$, $\sigma_\text{M}$ & \quad minimum and maximum values of the flat surface tension obtained from the simulations\\
    $\delta_\text{m}$, $\delta_\text{M}$ & \quad minimum and maximum values of the Tolman length obtained from the simulations\\
\end{tabular}\\
}

%
%
%

\bibliographystyle{ieeetr}
\bibliography{biblio_bibtex}
\end{document}